\documentclass[aps,prl,twocolumn,floatfix]{revtex4}% Physical Review Lett
\usepackage{amsmath}
\usepackage{url}
\usepackage{subfigure}
\usepackage{graphicx,psfrag}% Include figure files
\usepackage{bm}% bold math
\usepackage{multirow}

\newcommand{\nn}{\nonumber\\}
\newcommand{\ti}[1]{^{(#1)}}

\begin{document}

%\preprint{LAUR-08-????}

\title{When is social computation better than the sum of its parts?}
\author{Vadas Gintautas}
\email{vadasg@gmail.com}

\author{Aric Hagberg}
\email{hagberg@lanl.gov}

\author{Lu{\'i}s M. A. Bettencourt}
\email{lmbett@lanl.gov}

\affiliation{ Center for Nonlinear Studies, and Applied Mathematics and Plasma Physics, Theoretical Division, Los Alamos National Laboratory, Los Alamos NM 87545}
\date{\today}

\begin{abstract}
Social computation, whether in the form of searches performed by 
swarms of agents or collective predictions of markets, often supplies
remarkably good solutions to complex problems.
In many examples, individuals trying to solve a problem locally can aggregate their information and work together to arrive at a superior global solution. This suggests that there may be general principles of information aggregation and coordination that can transcend particular applications.  Here we show that the general structure of this problem can be cast in
terms of information theory and derive mathematical conditions that lead to optimal
multi-agent searches. Specifically, we illustrate the problem in terms of
local search algorithms for autonomous agents looking for the spatial
location of a stochastic source.  We explore the types of search
problems, defined in terms of the statistical properties of the source and the
nature of measurements at each agent, for which coordination among
multiple searchers yields an advantage beyond that gained by having
the same number of independent searchers.  We show that effective
coordination corresponds to synergy and that ineffective coordination
corresponds to independence as defined using information theory.  We
classify explicit types of sources in terms of their potential for
synergy.  We show that sources that emit uncorrelated signals 
provide no opportunity for synergetic
coordination while sources that emit signals that are correlated
in some way, do allow for strong synergy between searchers.  These general
considerations are crucial for designing optimal algorithms for
particular search problems in real world settings.
\end{abstract}

\maketitle

\section{Introduction}
The ability of agents to share information and to coordinate 
actions and decisions can provide significant practical
advantages in real-world searches. Whether the target is a person trapped by an avalanche or a hidden
cache of nuclear material, being able to deploy multiple autonomous
searchers can be more advantageous and safer than sending human operators.  For example, small autonomous, possibly expendable robots could
be utilized in harsh winter climates or on the battlefield.  

In some problems, e.g. locating a cellular
telephone via the signal strength at several towers, there
is often a simple geometrical search strategy, such as triangulation, which
works effectively. However, in search problems where the signal is stochastic or no geometrical solution is known, e.g. searching for a weak scent source in a
turbulent medium, new methods need to be developed.  This is especially true
when designing autonomous and self-repairing algorithms for robotic
agents~\cite{Bourgault_2003}.
Information theoretical methods provide a promising approach to
develop objective functions and search algorithms to fill this gap.
In a recent paper, Vergassola et al. 
demonstrated that infotaxis, which is motion based on expected
information gain, can be a more effective search strategy 
when the source signal is weak than conventional methods such as moving along the gradient of a chemical
concentration~\cite{Vergassola_2007}.  The infotaxis algorithm combines the two competing goals of 
exploration of possible search moves and exploitation of received signals to guide the searcher
in the direction with the highest probability of finding the source~\cite{Sulton_1998}.

To improve the efficiency of search by using more than
one searcher requires determining under what circumstances a collective
(parallel) search is better (faster) than the independent {\em
  combination} of the individual searches.  Much heuristic work,
anecdotally inspired by strategies in social insects and flocking
birds~\cite{couzin-2005-effective,bonabeau-2000-swarm}, 
has suggested that collective action should be
advantageous in searches in real world complex problems, such as
foraging, spatial mapping, and navigation. However, all approaches to
date rely on simple heuristics that fail to make explicit the
general informational advantages of such strategies.

The simplest extension of infotaxis to collective searches is to have multiple 
{\em  independent} (uncoordinated) searchers that share
information; this corresponds in general to a linear increase in
performance with the number of searchers.  However, given some general
knowledge about the structure of the search, substantial increases in
the search performance of a collective of agents can be achieved,
often leading to exponential reduction in the search effort, in terms
of time, energy or number of steps~\cite{Seung_1992,Freund_1997,
  Fine_2002}.  
In this work we explore how the concept of information synergy can be leveraged to 
improve infotaxis of multiple coordinated searchers. 
Synergy corresponds to the general situation when measuring two
or more variables {\em together} with respect to another (the target's signal) 
results in a greater information gain than the sum of that from each variable {\em separately}~\cite{Bettencourt_2007,  Bettencourt_2008}. 
We identify the types of spatial search problems for which coordination among multiple
searchers is effective (synergetic), as well as when it is ineffective, and corresponds to independence.  
We find that classes of statistical sources, such as those that emit uncorrelated signals (e.g. Poisson processes) 
provide no opportunity for synergetic coordination. On the other hand, sources that emit particles with spatial, 
temporal, or categorical correlations, do allow for strong synergy between searchers that can be exploited via coordinated motion. These considerations divide collective search problems into different general classes and are crucial for designing effective algorithms for particular applications.

\section{Information theory approach to stochastic search}
Effective and robust search methods for the location of stochastic sources must
balance the competing strategies of exploration and exploitation~\cite{Sulton_1998}.
On the one hand, searchers must exploit measured cues to guide their optimal next move. 
On the other hand, because this information is statistical, more measurements need to typically be made 
that are guided by different search scenarios.  Information theory approaches to search achieve this
balance by utilizing movement strategies that increase the expected
information gain, which in turn is a functional of the many possible source locations.  
In this section we define the necessary formalism and use it to set up the general structure 
of the stochastic search problem.

\subsection{Synergy and Redundancy}
First we define the concepts of information, synergy and redundancy
explicitly.   Consider the stochastic variables $X_i, i=1\ldots n$.  
Each variable $X_i$ can take on specific states, denoted by the
corresponding lowercase letter, that is $X$ can take on a set of states
$\{x\}$.  
For a single variable $X$ the Shannon entropy (henceforth ``entropy'') is $S(X) = -\sum_{x} P(x)\log_{2} P(x)$, where $P(x)$ is the probability that the variable $X$ take on the value $x$~\cite{Cover_1991}.  
The entropy is a measure of uncertainty about the state of $X$, therefore entropy can only decrease or remain unchanged as more variables are measured.  
The conditional entropy of a variable $X_1$ given a second variable $X_2$ is $S(X_{1}|X_{2}) = -\sum_{x_{1}, x_{2}} P(x_{1},x_{2})\log_{2} (P(x_{1},x_{2})/P(x_{2}))\leq S(X_{1})$.  
The mutual information between two variables, which plays an important role in search strategy, is defined as the change in entropy when a variable is measured $I(X_{1},X_{2}) = S(X_{1}) -  S(X_{1}|X_{2}) \geq 0$.
These definitions can be directly extended to multiple variables. For 3 variables, we make the following definition~\cite{Schneidman_2003}: $R(X_{1},X_{2},X_{3}) \equiv I(X_{1},X_{2}) - I(\{X_{1},X_{2}\}|X_{3})$.
This quantity measures the degree of ``overlap'' in the information contained in variables $X_{1}$ and $X_{2}$ with respect to $X_{3}$.  
If $R(X_{1},X_{2},X_{3}) > 0$, there is overlap and $X_{1}$ and $X_{2}$ are said to be redundant with respect to $X_{3}$.  
If $R(X_{1},X_{2},X_{3}) < 0$, more information is available when these variables are considered together than when considered separately.  
In this case $X_{1}$ and $X_{2}$ are said to be synergetic with respect to $X_{3}$.  
If $R(X_{1},X_{2},X_{3}) = 0$, $X_{1}$ and $X_{2}$ are independent~\cite{Bettencourt_2007, Bettencourt_2008}.

\subsection{Two-dimensional spatial search}
We now formulate the two-dimensional stochastic search problem.  We
consider, for simplicity, the case of two searchers seeking to find a
stochastic source located in a finite two-dimensional plane. 
This is a generalization of the 
single searcher formalism presented in Ref.~\cite{Vergassola_2007}.  
At any time step, the searchers have positions
$\{r_{i}\}, i=1,2$ and observe some number of particles $\{h_{i}\}$
from the source.  
The searchers do not get information about the trajectories or speed
of the particles; they only get information if a particle 
was observed or not.  Therefore
simple geometrical methods such as triangulation are not possible.
Let the variable $R_{0}$ correspond to all the possible locations of the
source $r_{0}$.  The searchers compute and share a probability
distribution $P\ti{t}(r_{0})$ for the source at each time index $t$.
Initially the probability for the source is assumed to be to be uniform.
After each measurement $\{h_{i},r_{i}\}$, the searchers update their estimated probability distribution of source positions via Bayesian inference.  
First the conditional probability $P\ti{t+1}(r_{0}|\{h_{i},r_{i}\}) \equiv P\ti{t}(r_{0})P(\{h_{i},r_{i}\}|r_{0})/A$, is calculated, where $A$ is a normalization
over all possible source locations as required by Bayesian inference.  This is then assimilated via Bayesian update so that $P\ti{t+1}(r_{0}) \equiv P\ti{t+1}(r_{0}|\{h_{i},r_{i}\})$.

If the searchers do not find the source at their present
locations they choose the next local move using an infotaxis step
to maximize the expected information gain.
To describe the infotaxis step we first need some definitions.
The entropy of the distribution $P\ti{t}(r_{0})$ at time $t$ is
defined as
 $S\ti{t}(R_{0}) \equiv -\sum_{r_{0}} P\ti{t}(r_{0}) \log_{2} P\ti{t}(r_{0})$.
In terms of a specific measurement $\{h_{i},r_{i}\}$
the entropy is ({\em before} the Bayesian update)
 $S\ti{t}_{\{h_{i},r_{i}\}}(R_{0}) \equiv -\sum_{r_{0}}P\ti{t}(r_{0}|\{h_{i},r_{i}\})\log_{2} P\ti{t}(r_{0}|\{h_{i},r_{i}\})$.
We define the difference between the entropy at time $t$ and the 
entropy at time $t+1$ after a measurement $\{h_{i},r_{i}\}$ to be 
 $\Delta S\ti{t+1}_{\{h_{i},r_{i}\}} \equiv S\ti{t+1}_{\{h_{i},r_{i}\}}(R_{0}) - S\ti{t}(R_{0})$.

 Initially the entropy is at its maximum for a uniform prior: $ S\ti{0}(R_{0})=\log_{2} N_{s}$, where $N_{s}$ is the number of possible locations for the source in a discrete space.
For each possible joint move $\{r_{i}\}$, the change in
expected entropy $\overline{\Delta S}$ is computed and the move with the
minimum (most negative) $\overline{\Delta S}$ is executed.
The expected entropy is computed by considering the reduction
in entropy for all of the possible joint moves

\begin{align}
    \overline{\Delta S} &= -\biggl[\sum_{i}P\ti{t}(R_{0}=r_{i})\biggr] S\ti{t}(R_{0})\nn
    &+ \biggl[1 - \sum_{i}P\ti{t}(R_{0}=r_{i})\biggr]
    \Delta S\ti{t+1}_{\{h_{i},r_{i}\}}\nn
    &\times \sum_{ h_{1}, h_{2} }
     \biggl[\sum_{r_{0}} P\ti{t}(r_{0})P\ti{t+1}(\{h_{i},r_{i}\}|r_{0})\biggr].
\label{eq:deltasbar}
\end{align}

The first term in Eq.~\eqref{eq:deltasbar} corresponds to one of the
searchers finding the source in the next time step (the final entropy will be $S=0$ so $\overline{\Delta S}=-S$).
The second term considers the reduction in entropy for all possible measurements
at the proposed location, weighted by the probability of each of those
measurements.  The probability of the searchers obtaining the
measurement $\{h_{i}\}$ at the location $\{r_{i}\}$ is given by the
trace of the probability $P\ti{t+1}(\{h_{i},r_{i}\}|r_{0})$ over all
possible source locations.

\section{Correlated stochastic source and synergy of searchers}

The expected entropy reduction $\overline{\Delta S}$ is calculated for {\em joint} moves of the
searchers, that is, all possible combinations of individual moves.
Compared with multiple independent searchers this calculation incurs some extra computational cost.
Thus, when designing a search algorithm, it is important to
know whether an advantage (synergy) can be gained by considering joint moves
instead of individual moves.   
Since the search is based on optimizing the maximum information gain we 
need to explore if joint moves are synergetic or redundant.
In this section we will show how correlations in the source
affect the synergy and redundancy of the search.

\begin{figure*}[htb]
\begin{center}
    \includegraphics[width=1.42\columnwidth]{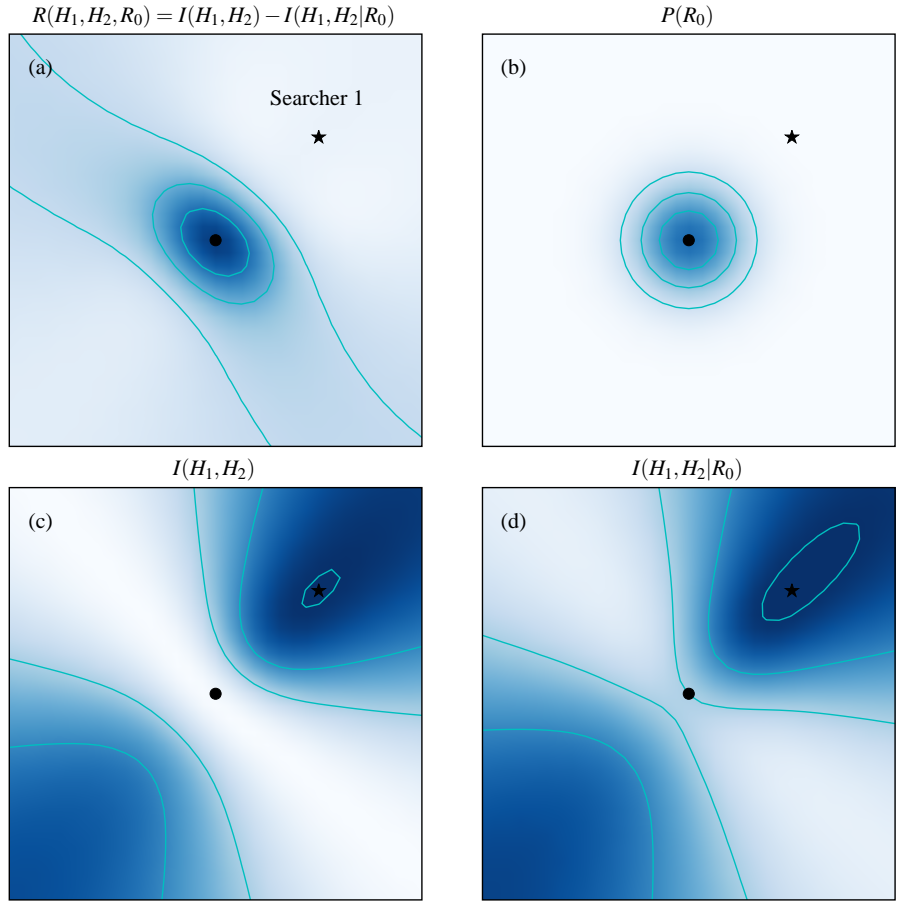}
\end{center}
\caption{Synergy for the two searcher problem with angular correlations.  (a)
  $R(H_{1},H_{2},R_{0})$ as a function of the position
  of searcher 2  ($r_{2}$) for a fixed location of searcher 1 
  ($r_1$, shown as a black star).
  The most probable source location is in the center (black dot).
  The white to blue scale indicates $R=0$ to $R=-2\times 10^{-5}$
  and we note that $R \leq 0$ everywhere.
  The darker color indicates stronger
  synergy values when searcher 2 is near the source.  The synergy is 
  less when searcher 2 is away from or on the opposite side
  ($R\approx0$) of the source.
  (b) The probability distribution of source locations, peaked at the
  center: $P(\vec{r_{0}}) = A\exp{(-|\vec{r_{0}}|^{2}/0.02)}$, where 
  $A=1/\sum_{\vec{r_{0}}} P(\vec{r_{0}})$ is a normalization factor.
  White to blue indicates $P=0$ to $P=0.02$.
  (c) $I(H_{1},H_{2})$; (d) $I(H_{1},H_{2}|R_{0})$.
  In (c) and (d) white to blue indicates $I=0$ to $I=6\times10^{-5}$.
  Contour lines have been added to guide the eye.
  In all frames the data is plotted in 
  a two-dimensional spatial domain of $x,y=[-0.5,0.5]$
%  The horizontal and vertical axes in all plots 
%  range from $-\tfrac{1}{2}$ to $\tfrac{1}{2}$
  and all vectors are measured from the origin $x=0,y=0$.
  The parameter $\sigma^{2} = 1.1$ in  Eq.~\ref{eq:fdefn}. 
  }
\label{fig:Rh1h2r0}
\end{figure*}

In the following we will assume there are no radial correlations between
particles emitted from the source and that the probability of detecting particles decays with distance to the source. 
For each particle emitted from the source, the searcher $i$
has an associated actual probability $\pi_{i}(r_{0})$ of catching the
particle.  
The probability $\pi_{i}(r_{0})$ is defined in terms of a possible source
location $r_{0}$ and  the location $r_{i}$ of searcher $i$: 
$\pi_{i}(r_{0}) = B\exp{(-|\vec{r}_{i}-\vec{r_{0}}|^2)}$, where $\{r_{i}\}$ is the set of all the
searcher positions and $B$ is a normalization constant.  Note that this is just the
radial component of the probability; if there are angular correlations
these are treated separately.  We may now write $R$, as a function of
the variables $R_{0}$, $H_{1}$, and $H_{2}$, in terms of the
conditional probabilities:
\begin{align}
    &R(H_{1},H_{2},R_{0}) =\nn
    &\sum_{h_{1},h_{2},r_{0}} P(r_{0},h_{1},h_{2}) \log_{2} \frac{ P(h_{1}| r_{0}) P(h_{2}| r_{0}) P(h_{2}|h_{1}) }{ P(h_{2}) P(h_{1},h_{2}|r_{0})}.
\label{eq:Rcondprobs}
\end{align}
It is sufficient for $R \neq 0$ that the argument of the logarithm differs from $1$. 
This can be achieved even if measurements are conditionally independent (redundancy), mutually independent (synergy), or when neither of these conditions apply.  

\subsection{Uncorrelated signals: Poisson source}
\label{sec:poisson}
First, consider a source which emits particles according to a Poisson
process with known mean $\lambda_{0}$ so emitted particles are completely uncorrelated spatially and temporally.  
%There is
%no bound on the number of particles emitted by the source, so there is
%a combinatorial possible number of hits at each searcher.  
If searcher
1 is able to get a particle that has already been detected by searcher
2, it is clear that the searchers are completely independent and there
is no chance of synergy.  It may appear at first that
implementing a simple exclusion where two searchers cannot get the
same particle would be enough to foster cooperation between searchers.
We will instead show that it is the Poisson nature of the source that
makes synergy impossible, even under mutual exclusion of the
measurements.  

The probability of the measurement
$\{h_{i}\}$ is given by
\begin{equation}
    P(\{h_{i},r_{i}\}|r_{0}) = \sum_{h_{s}=\sum_{i}h_{i}}^{\infty} P_{0}(h_{s},\lambda_{0}) M(\{\pi_{i}(r_{0})\},\{h_{i}\},h_{s}).
    \label{eq:Pmeasgivenr0}
\end{equation}
The sum is over all possible values of $h_{s}$, weighted by the
Poisson probability mass function with the known mean $\lambda_{0}$.
$M$ is the probability mass function of the multinomial distribution
for that measurement; it handles the combinatorial degeneracy and the exclusion.
It is not difficult to show by summing over $h_{s}$ that
$P(\{h_{i},r_{i}\}|r_{0})$ can be written as a product of Poisson
distributions with effective means $\lambda_{0}\pi_{i}$,
\begin{equation}
    P(\{h_{i},r_{i}\}|r_{0}) =
    \frac{\lambda_{0}^{\sum_{i}h_{i}} e^{-\lambda_{0}\sum_{i}\pi_{i}} \prod_{i}\pi_{i}^{h_{i}}}{\prod_{i}(h_{i}!)}
     = \prod_{i}P_{0}(h_{i},\lambda_{0}\pi_{i}).
     \label{eq:factors}
\end{equation}
At this point we consider whether a search like this can be synergetic
for the 2 searcher case.  Eq.~\eqref{eq:factors} shows that the two measurements are conditionally independent and therefore $P(h_{1},h_{2}|r_{0}) = P(h_{1}|r_{0}) P(h_{2}|r_{0})$.
It follows from Eq.~\eqref{eq:Rcondprobs} that $R(H_{1},H_{2},R_{0})$ $= I(H_{1},H_{2}) \geq 0$.
Therefore the searchers are either redundant (if the measurements interfere with each other) or independent with respect to the source. Synergy is impossible so that searchers gain no advantage by considering
joint moves. The only advantage of coordination comes possibly from avoiding positions that lead to a decrease in performance of the collective due to competition for the same signal.

\subsection{Correlated signals: angular biases}
\label{sec:angular}
We now consider a source that emits particles that are spatially
correlated.  We assume for simplicity that at each time step the source emits 2
particles.  The first particle is emitted at a random angle
$\theta_{h_{1}}$ chosen uniformly from $[0,2\pi)$.  The second
particle is emitted at an angle $\theta_{h_{2}}$ with probability 
\begin{equation}
P(\theta_{h_{2}}|\theta_{h_{1}}) = D\exp{[-(|\theta_{h_{1}}-\theta_{h_{2}}|-\pi)^{2}/\sigma^{2}]} \equiv f,
\label{eq:fdefn}
\end{equation}
where $D$ is a normalization factor.
The searchers are assumed to know the variance $\sigma$ for
simplicity; this is a reasonable assumption if the searchers have any
information about the nature of the target (just as for the Poisson
source they had statistical knowledge of the parameter $\lambda_{0}$).
The calculation of the conditional probability
$P(\{h_{i},r_{i}\}|r_{0})$ requires some care.  Specifically, this
quantity is the probability of the measurement $\{h_{i}\}$, assuming a
certain source position.  Since there are 2 particles emitted at each
time step, there are 4 possible cases, each with a different
probability, as shown in Table~\ref{tab:pth1th2}.  Here
$\theta_{h_{1}}$ and $\theta_{h_{2}}$ are calculated from $r_{1}$ and
$r_{2}$, respectively: $\theta_{h_{i}}\equiv
\arctan{\frac{r_{0,y}-r_{i,y}}{r_{0,x}-r_{i,x}}}$.  Note that the
$\pi_{i}$ are functions of $r_{0}$. The coefficient $D$ is chosen such that
$\tfrac{1}{N}\sum_{h_2}\sum_{r_2} P(\{h_{1},h_{2},r_{1},r_{2}\}|r_{0}) = P(\{h_{1},r_{1}\}|r_{0})$, corresponding to the normalization condition $\tfrac{1}{N}\sum_{r_2} f = 1$.

\begin{table*}
\begin{center}
    \begin{tabular}{|c|c|c|c|c|}
    \hline
$\{h_{1},h_{2}\}$ &$ \{1,1\}$ &$ \{1,0\}$ &$ \{0,1\}$ & $\{0,0\}$ \\
\hline
$P(\{h_{1},r_{1}\}|r_{0})  $&$ \pi_{1} $&$ \pi_{1} $&$1-\pi_{1} $&$ 1-\pi_{1}$\\
$P(\{h_{2},r_{2}\}|r_{0})  $&$ \pi_{2} $&$ \pi_{2} $&$1-\pi_{2}  $&$ 1-\pi_{2}$\\
$P(\{h_{1},h_{2},r_{1},r_{2}\}|r_{0}) $&$ \pi_{1}\pi_{2}f^{2} $ &$ \pi_{1}f\bigl(1-\pi_{2}f\bigr) $&$ \pi_{2}f\bigl(1-\pi_{1}f\bigr) $&$ (1-\pi_{1}f)(1-\pi_{2}f)$ \\
\hline
\end{tabular}
\end{center}
\caption{Probability calculation for all possible states in the
  correlated source search. 
  Here $\pi_{i}(r_{0})=B\exp{(-|\vec{r}_{i}-\vec{r_{0}}|^2)}$ 
  is written as $\pi_{i}$ to save space.}
\label{tab:pth1th2}  
\end{table*}

Figure~\ref{fig:Rh1h2r0} shows the value of $R(H_{1},H_{2},R_{0})$
and the values of the mutual informations $I(H_1,H_2)$ and
$I(H_1,H_2|R_0)$ for each possible position $r_2$ of searcher 2.
We assume a nonuniform, peaked probability distribution
for the source [Figure~\ref{fig:Rh1h2r0}(b)] and that
the position of searcher 1 is fixed. 
In this setup we see that $R<=0$ for every possible position of
searcher 2 indicating that only synergy is possible.
This is a consequence of the angular spatial correlation between the
particles emitted by the source.
The synergy is highest near the source location, where the
source probability is strongly peaked, and falls off rapidly
away from the source location.
Furthermore there is little to no
synergy near searcher 1 since in that region it is very unlikely that
both searchers would simultaneously observe a particle.  The area of
greatest synergy corresponds to the most probable source locations 
for both searchers to simultaneously observe a particle.
$P(r_{0})$ is very flat at the boundaries; thus $R_{0}$ contributes
little to $I(H_{1},H_{2}|R_{0})$ in the lower left corner and $R$ is
small.

\section{Conclusion}
In the real world, communication between agents, as well as
centralized or decentralized real-time computation can be difficult or
expensive.  Therefore it is important to consider the classes of search
problems for which coordination between searchers can achieve quantitative advantages over independent agents.  
In this work we studied search algorithms for autonomous
agents looking for the spatial location of a stochastic source.  We
defined the search problem for multiple agents in terms of infotaxis
[see Eq.~\eqref{eq:deltasbar}].  We also showed why synergy gives rise
to an advantage in this type of search.  We considered two types of
sources. We first demonstrated that a source emitting uncorrelated
particles will afford no opportunity for synergy (see
  Section~\ref{sec:poisson}).  In a search for a Poisson source,
multiple coordinated searchers (ones that consider sets of joint moves
rather than each considering an independent move) can not hope to do
better than multiple independent searchers.  Next we showed that, for
a source emitting particles with (angular) correlations (see Section~\ref{sec:angular}),
only synergy or independence is possible (see Fig.~\ref{fig:Rh1h2r0}).  The ability of the searchers
to leverage synergy depends strongly on their ability to estimate with
some accuracy the probability distribution of source locations.  These
general considerations are crucial for the exploitation of social computation in terms of the design of 
optimal collective algorithms in particular applications.
The next step to making this approach applicable to a broader class of 
problems, including those not limited to spatial searches, is to generalize the results 
to more than 2 searchers and to explore how synergy may be best 
leveraged to give increases in search speed and efficiency.

This paper is released under LA-UR 09-00432.
%\bibliographystyle{unsrt}
%\bibliography{synergy}

\end{document}